\documentclass[prb,aps,preprint,eqsecnum]{revtex4}

\usepackage[centertags]{amsmath}
\usepackage{amssymb}
\usepackage{amsthm}

\begin{document}

\title{Anharmonic quantum contribution to vibrational dephasing}
\author{Debashis~Barik and Deb~Shankar~Ray{\footnote{Email address:
pcdsr@mahendra.iacs.res.in}}} \affiliation{Indian Association for
the Cultivation of Science, Jadavpur, Kolkata 700 032, India}


\begin{abstract}
Based on a quantum Langevin equation and its corresponding
Hamiltonian within a c-number formalism we calculate the
vibrational dephasing rate of a cubic oscillator. It is shown
that leading order quantum correction due to anharmonicity of the
potential makes a significant contribution to the rate and the
frequency shift. We compare our theoretical estimates with those
obtained from experiments for small diatomics $N_2$, $O_2$ and
$CO$.
\end{abstract}


\maketitle

\section{{\bf Introduction}}
 A molecule in a liquid undergoes random frequency
fluctuations due to the stochastic forces imparted by the
surrounding medium. The correlation between this frequency
fluctuations results in vibrational dephasing. The problem has
received wide attention both from theoretical and experimental
point of view over the last couple of decades. Several approaches
to understand the basic nature of vibrational dephasing have been
made. These include notably binary collision theory \cite{fis} in
which fluctuations are taken into account in terms of collisional
events, hydrodynamic model \cite{ox1,ox2,ox3,geo} relating random
force and shear viscosity of the liquid, Hamiltonian theory
\cite{pol} in terms of normal mode analysis and numerical
simulations \cite{ox4,bj,yan,bag1,bag2} using various molecular
potentials. A key element of these formulations is the realization that
vibrational dephasing owes its origin to cubic anharmonicity of
the vibrational potential. In the present paper we attempt to
explore further this issue within a quantum mechanical content.

A good number of approaches to vibrational dephasing make use of
generalized Langevin equation that governs the dynamics of the
system by an infinite number of harmonic oscillators coupled
linearly to the system. Very recently based on a coherent state
representation of noise operator and a positive definite Wigner
canonical thermal distribution \cite{hil} of bath oscillators a
c-number quantum Langevin equation
\cite{db1,skb,db2,db3,bk1,bk2,db4} in the context of rate theory
and stochastic processes has been proposed. An important offshoot
of this formulation is that it takes care of quantum correction
due to nonlinearity of the system potential order by order. It
would thus seem that one should be able to analyze the vibrational
dephasing rate quantum mechanically for arbitrary noise
correlation and temperature and explore the role of this quantum
correction to anharmonicity. This is the main purpose of this
paper. In what follows we construct a c-number Hamiltonian within
a normal mode description and estimate the dephasing rate by
calculating the effective correlation between the relevant
c-number co-ordinates of anharmonic origin. Since for a small
diatomic molecule like $N_2$, the fundamental vibrational
frequency $\omega_0$ is so high ($2326 \;cm^{-1}$) that the ratio
$\hbar \omega_0/{k T}$ is as large as $43.52$ even at, say, $77\;
^0 K$, it is imperative that quantum nature of the oscillator
molecule is significant in estimating the dephasing rate in the
harmonic as well as in the anharmonic regime. With this end in
view we examine the vibrational dephasing rate to estimate the
anharmonic quantum correction to this rate and its variation with
temperature away from critical or triple point for three widely
studied diatomics $N_2$, $O_2$ and $CO$ to allow ourselves a fair
comparison with experiments \cite{wrl,sco,lau,srj,clo1,clo2,jen}.

The outlay of the paper is as follows: In Sec.II we introduce the
quantum Langevin equation and its Hamiltonian counterpart within
a c-number normal mode description. Since the cubic nonlinearity
gives rise to a leading order contribution to dephasing rate we
estimate the quantum vibrational dephasing rate for a cubic
oscillator in Sec.III. The quantum corrections due to
nonlinearity of the system potential is calculated explicitly in
Sec.IV. Sec.V is devoted to the results obtained theoretically
for three diatomics $N_2$, $O_2$ and $CO$ which are compared with
experiments. The paper is concluded in Sec. VI.

\section{{\bf C-number quantum Langevin equation and normal mode transformation}}

We consider a particle of mass $\mu$ coupled to a medium comprised
of a set of harmonic oscillators with frequency $\omega_i$. This
is described by the following Hamiltonian \cite{pol}:

\begin{equation}\label{1}
H = \frac{\hat{p}^2}{2\mu} + V(\hat{q}) + \sum^N_{i=1} \left \{
\frac{\hat{p}_i^2}{2m_i} + \frac{m_i}{2} (\omega_i \hat{x}_i +
\frac{c_i}{m_i \omega_i} \hat{q} )^2 \right \}
\end{equation}

Here $\hat{q}$ and $\hat{p}$ are co-ordinate and momentum
operators of the particle and the set $\{ \hat{x}_i,\hat{p}_i \}$
is the set of co-ordinate and momentum operators for the
reservoir oscillators of mass $m_i$ coupled linearly to the system
through their coupling coefficients $c_i$. The potential
$V(\hat{q})$ is due to the external force field for the Brownian
particle. The co-ordinate and momentum operators follow the usual
commutation relations [$\hat{q}, \hat{p}$] = $i \hbar$ and
[$\hat{x}_i, \hat{p}_j$] = $i \hbar \delta_{ij}$.

Eliminating the reservoir degrees of freedom in the usual way we
obtain the operator Langevin equation for the particle,

\begin{equation}\label{2}
\mu \ddot{ \hat{q} } (t) + \mu \int_0^t dt'  \gamma (t-t') \dot{
\hat{q} } (t') + V' ( \hat{q} ) = \hat{F} (t) \; \; ,
\end{equation}

where the noise operator $\hat{F} (t)$ and the memory kernel
$\gamma (t)$ are given by

\begin{equation}\label{3}
\hat{F} (t) = - \sum_j \left [ \left \{ \frac{m_j \omega_j^2}{c_j}
\hat{x}_j (0) + \hat{q} (0) \right \} \frac{c_j^2}{m_j \omega_j^2}
\cos \omega_j t + \frac{c_j}{m_j \omega_j} \hat{p}_j (0) \sin
\omega_j t \right ]
\end{equation}

and

\begin{equation}\label{4}
\gamma (t) = \sum_{j=1}^N \frac{c_j^2}{m_j \omega_j^2} \cos
\omega_j t \; \; ,
\end{equation}

The Eq.(\ref{2}) is the well known exact quantized operator
Langevin equation for which the noise properties of $\hat{F} (t)$
can be derived by using a suitable initial canonical distribution
of the bath co-ordinate and momentum operators at $t=0$ as
follows;

\begin{eqnarray}
\langle \hat{F} (t) \rangle_{QS} &=& 0 \label{5} \\
\frac{1}{2} \{ \langle\hat{F}(t)\hat{F}(t^\prime)\rangle_{QS} +
\langle\hat{F}(t^\prime)\hat{F}(t)\rangle_{QS} \} &=& \frac{1}{2}
\sum_{j=1}^N \frac{c_j^2}{m_j \omega_j^2} \hbar \omega_j
\left(\coth \frac{\hbar\omega_j}{2 k_B T}\right) \cos
\omega_j(t-t^\prime)\label{6}
\end{eqnarray}

where $\langle...\rangle_{QS}$ refers to quantum statistical
average on bath degrees of freedom and is defined as

\begin{equation}\label{7}
\langle \hat{O} \rangle_{QS} = \frac{{\rm Tr} \hat{O} \exp
(-\hat{H}_{{\rm bath}}/k_BT)}{{\rm Tr}\exp (-\hat{H}_{{\rm
bath}}/k_BT)}
\end{equation}

for any operator $\hat{O}(\{(m_j \omega_j^2/c_j)\hat{x}_j +
\hat{q}\},\{\hat{p}_j\})$ where $\hat{H}_{{\rm bath}}
(\sum^N_{i=1} (\hat{p}_i^2/{2m_i} + m_i/2 (\omega_i \hat{x}_i +
\frac{c_i}{m_i\omega_i} \hat{q} )^2))$ at $t=0$. By Trace we mean
the usual quantum statistical average. Eq.(\ref{6}) is the
fluctuation-dissipation relation with the noise operators ordered
appropriately in the quantum mechanical sense.

To construct a c-number Langevin equation
\cite{db1,skb,db2,db3,bk1,bk2,db4} we proceed from Eq.(\ref{2}).
We first carry out the {\it quantum mechanical average} of
Eq.(\ref{2})

\begin{equation}\label{8}
\mu \langle \ddot{ \hat{q} } (t) \rangle + \mu \int_0^t dt' \gamma
(t-t') \langle \dot{ \hat{q} } (t') \rangle + \langle V' (
\hat{q} ) \rangle = \langle \hat{F} (t) \rangle
\end{equation}

where the quantum mechanical average $\langle \ldots \rangle$ is
taken over the initial product separable quantum states of the
particle and the bath oscillators at $t=0$, $| \phi \rangle \{ |
\alpha_1 \rangle | \alpha_2 \rangle \ldots | \alpha_N \rangle \}
$. Here $| \phi \rangle$ denotes any arbitrary initial state of
the particle and $| \alpha_i \rangle$ corresponds to the initial
coherent state of the $i$-th bath oscillator. $|\alpha_i \rangle$
is given by $|\alpha_i \rangle = \exp(-|\alpha_i|^2/2)
\sum_{n_i=0}^\infty (\alpha_i^{n_i} /\sqrt{n_i !} ) | n_i \rangle
$, $\alpha_i$ being expressed in terms of the mean values of the
shifted co-ordinate and momentum of the $i$-th oscillator,
$\{(m_i\omega_i^2/c_i)\langle \hat{x}_i (0) \rangle + \langle
\hat{q}(0) \rangle\} =  \sqrt{\hbar /2m_i\omega_i} (\alpha_i +
\alpha_i^\star )$ and $\langle \hat{p}_i (0) \rangle = i
\sqrt{\hbar m_i \omega_i/2 } (\alpha_i^\star - \alpha_i )$,
respectively. It is important to note that $\langle \hat{F} (t)
\rangle$ of Eq.(\ref{8}) is a classical-like noise term which, in
general, is a non-zero number because of the quantum mechanical
averaging and is given by $(\langle \hat{F}(t) \rangle \equiv
f(t))$;

\begin{equation}\label{9}
f(t) = - \sum_j \left [ \left \{ \frac{m_j \omega_j^2}{c_j}
\langle \hat{x}_j (0)\rangle + \langle\hat{q} (0)\rangle \right \}
\frac{c_j^2}{m_j \omega_j^2} \cos \omega_j t + \frac{c_j}{m_j
\omega_j} \langle\hat{p}_j (0)\rangle \sin \omega_j t \right ]
\end{equation}

It is convenient to rewrite the $c$-number equation (\ref{8}) as
follows;

\begin{equation}\label{10}
\mu \langle \ddot{ \hat{q} } (t) \rangle + \mu \int_0^t dt' \gamma
(t-t') \langle \dot{ \hat{q} } (t') \rangle + \langle V' (
\hat{q} ) \rangle = f (t)
\end{equation}

To realize $f(t)$ as an effective c-number noise we now introduce
the ansatz \cite{hil,db1,skb,db2,db3,bk1,bk2,db4} that the momenta
$\langle \hat{p}_j (0) \rangle$ and the shifted co-ordinates
$\{(m_j \omega_j^2/c_j)\langle\hat{x}_j(0)\rangle +
\langle\hat{q}(0)\rangle\}$, $\{\hat{p}_j\}$ of the bath
oscillators are distributed according to a canonical distribution
of Gaussian form as

\begin{equation}\label{11}
{\cal P}_j = {\cal N} \exp \left \{ - \frac{ [ \langle \hat{p}_j
(0) \rangle^2 + \frac{c_j^2}{\omega_j^2} \{ \frac{m_j
\omega_j^2}{c_j}\langle \hat{x}_j (0) \rangle + \langle \hat{q}
(0) \rangle \}^2 ] }{ 2 \hbar m_j \omega_j \left (
\bar{n}_j(\omega_j) + \frac{1}{2} \right ) } \right \}
\end{equation}

so that for any function of the quantum mechanical mean values of
the bath operator  $O_j ( \langle\hat{p}_j (0) \rangle, (({m_j
\omega_j^2}/{c_j})\langle\hat{x}_j (0) \rangle  + \langle \hat{q}
(0) \rangle ))$ the statistical average $\langle \ldots
\rangle_S$ is

\begin{eqnarray}
\langle O_j \rangle_S & = & \int O_j \;{\cal P}_j \; d\langle
\hat{p}_j (0) \rangle \;d \{ (m_j \omega_j^2/c_j)\langle \hat{x}_j
(0) \rangle + \langle \hat{q} (0) \rangle \} \; \; . \label{12}
\end{eqnarray}

\noindent Here $\bar{n}_j$ indicates the average thermal photon
number of the $j$-th oscillator at temperature $T$ and
$\bar{n}_j(\omega_j) = 1/[\exp \left ( \hbar \omega_j/k_BT \right
) - 1]$ and ${\cal N}$ is the normalization constant.

The distribution (\ref{11}) and the definition of statistical
average (\ref{12}) imply that $f(t)$ must satisfy

\begin{equation}\label{13}
\langle f(t) \rangle_S = 0
\end{equation}

and

\begin{equation}\label{14}
\langle f(t)f(t^{\prime})\rangle_S = \frac{1}{2} \sum_j
\frac{c_j^2}{m_j \omega_j^2}\hbar \omega_j \left ( \coth \frac {
\hbar \omega_j }{ 2k_BT } \right ) \cos \omega_j (t - t^{\prime})
\end{equation}

\noindent That is, $c$-number noise $f(t)$ is such that it is
zero-centered and satisfies the standard fluctuation-dissipation
relation (FDR) as expressed in Eq.(\ref{6}). It is important to
emphasize that the ansatz (\ref{11}) is a canonical thermal Wigner
distribution \cite{hil} for a shifted harmonic oscillator which
remains always a positive definite function. A special advantage
of using this distribution is that it remains valid as pure state
non-singular distribution function at $T = 0$. Furthermore, this
procedure allows us to {\it bypass the operator ordering}
prescription of Eq.(\ref{6}) for deriving the noise properties of
the bath in terms of fluctuation-dissipation relation and to
identify $f(t)$ as a classical looking noise with quantum
mechanical content. The procedure has been used by us in several
recent contexts \cite{db1,skb,db2,db3,bk1,bk2,db4}.

We now return to Eq.(\ref{10}) to add the force term
$V^\prime(\langle\hat{q} \rangle)$ on both sides of Eq.(\ref{10})
and rearrange it to obtain

\begin{eqnarray}
\mu \dot q &=& p
\label{15}\\
\dot p &=& - \int_0^t dt^\prime \gamma (t - t^\prime) p(t^\prime)
- V^\prime (q) + f(t) + Q(t) \label{16}
\end{eqnarray}

where we put $\langle\hat{q}(t)\rangle = q(t)$ and
$\langle\hat{p}(t)=p(t)$ for notational convenience and

\begin{equation}\label{17}
Q(t) = V^\prime(\langle\hat{q}\rangle) - \langle V^\prime(\hat{q})
\rangle
\end{equation}

represents the quantum correction due to the system degrees of
freedom. Eq.(\ref{16}) offers a simple interpretation. This
implies that the quantum Langevin equation is governed by a
$c$-number quantum noise $f(t)$ originating from the heat bath
characterized by the properties (\ref{13}) and (\ref{14}) and a
quantum fluctuation term $Q(t)$ characteristic of the
non-linearity of the potential.

Referring to the quantum nature of the system in the Heisenberg
picture, one may write.

\begin{eqnarray}
\hat{q} (t) &=& q + \delta\hat{q} \label{18}  \\
 \hat{p} (t) &=& p + \delta\hat{p}\label{19}
\end{eqnarray}

where $\langle\hat{q}\rangle(=q)$ and $\langle\hat{p}\rangle(=p)$
are the quantum-mechanical averages and $\delta\hat{q}$,
$\delta\hat{p}$ are the operators. By construction
$\langle\delta\hat{q}\rangle$ and $\langle\delta\hat{p}\rangle$
are zero and $[\delta\hat{q},\delta\hat{p}] = i\hbar$. Using
Eqs.(\ref{18}) and (\ref{19}) in $\langle V^{\prime} (\hat{q})
\rangle$ and a Taylor series expansion around
$\langle\hat{q}\rangle$ it is possible to express $Q(t)$ as

\begin{equation}\label{20}
Q(t) = -\sum_{n \ge 2} \frac{1}{n!} V^{(n+1)} (q)
\langle\delta\hat{q}^n(t)\rangle
\end{equation}

Here $V^{(n)}(q)$ is the n-th derivative of the potential $V(q)$.
For example, the second order $Q(t)$ is given by $Q(t) =
-\frac{1}{2} V^{\prime\prime\prime} (q) \langle \delta \hat{q}^2
\rangle$. The calculation of $Q(t)$
\cite{db1,skb,db2,db3,bk1,bk2,db4,sm,akp} therefore rests on
quantum correction terms, $\langle \delta \hat{q}^n (t) \rangle
(=B_n(t))$ which are determined by solving the quantum correction
equations as discussed in the Sec.IV.

The c-number Hamiltonian corresponding to Langevin equation
(\ref{15}, \ref{16}) is given by

\begin{equation}\label{21}
H = \frac{p^2}{2 \mu} + \left[ V(q) + \sum_{n \geq 2} \frac{1}{n!}
V^{(n)}(q) B_n(t) \right] + \sum^N_{i=1} \left \{ \frac{p_i^2}{2
m_i} + \frac{m_i}{2} ( \omega_i x_i + \frac{c_i}{m_i \omega_i} q
)^2 \right \}
\end{equation}

Note that the above Hamiltonian is different from our starting
Hamiltonian operator (\ref{1}) because of the c-number nature of
(\ref{21}). $\{x_i,p_i\}$ are the quantum mean values of the
co-ordinate and the momentum operators of the bath oscillators.

To characterize the properties of the bath we define, as usual,
the spectral density function as

\begin{equation}\label{22}
J(\omega)=\frac{\pi}{2 \mu} \sum_{i=1}^N \frac{c_i^2}{m_i
\omega_i}\; \delta(\omega-\omega_i)
\end{equation}

Splitting the potential into a linear and nonlinear part as

\begin{equation}\label{23}
V(q)=\frac{1}{2} \mu \omega_0^2 q^2 + V_1(q)
\end{equation}

where $V_1(q)$ is the nonlinear part of the classical potential
$V(q)$ we express, using (\ref{23}), the quantum correction term
as

\begin{eqnarray}\label{24}
\sum_{n\geq 2}\frac{1}{n!} V^{(n)}(q) B_n(t) & = & \frac{1}{2} \mu
B_2(t)\omega_0^2 +V_2(q)\\
V_2(q)& = & \sum_{n\geq 2}\frac{1}{n!} B_n (t)
V_1^{(n)}(q)\nonumber
\end{eqnarray}

Therefore the c-number Hamiltonian (Eq. 2.21) can be written as

\begin{equation}\label{25}
H=H_0+V_N(q)
\end{equation}

where, $H_0$, the harmonic part of the Hamiltonian is given by

\begin{equation}\label{26}
H_0 = \frac{p^2}{2 \mu} + \frac{1}{2}\mu\; \omega_0^2 q^2 +
\frac{1}{2} \mu B_2 \omega_0^2 + \sum^N_{i=1} \left \{
\frac{p_i^2}{2 m_i} + \frac{m_i}{2} ( \omega_i x_i +
\frac{c_i}{m_i \omega_i} q )^2 \right \}
\end{equation}

and $V_N(q)$ is given by

\begin{equation}\label{27}
V_N(q)=V_1(q)+V_2(q)
\end{equation}

$V_2(q)$ is the quantum correction term due to nonlinear part of
the system potential. We define the mass weighted co-ordinates as

\begin{equation}\label{28}
q^{\prime} = \sqrt{\mu}\; q \;\;\; and \;\;\;
x_j^{\prime}=\sqrt{m_j}\; x_j
\end{equation}

Diagonalizing the force constant matrix $T$ of the Hamiltonian
(Eq.2.26) as

\begin{equation}\label{29}
U\;T=\lambda^2\;U
\end{equation}

where $U$ provides the transformation from old co-ordinates to
the normal co-ordinates \cite{pol,p1}

\begin{eqnarray}
\left(
\begin{array}{cc}
{\rho}\\
{y_1}\\
{y_2}\\
{.}\\
{.}\\
{y_N}
\end{array}
\right) = U \left(
\begin{array}{cc}
{q^{\prime}}\\
{x_1^{\prime}}\\
{x_2^{\prime}}\\
{.}\\
{.}\\
{x_N^{\prime}}
\end{array}
\right)\label{30}
\end{eqnarray}

Here $\rho$ and $y_i$ are normal co-ordinates corresponding to
system and bath modes respectively.

The normal mode Hamiltonian for the harmonic part $H_0$ is then
given by \cite{pol},

\begin{equation}\label{31}
H_{NM} = \frac{1}{2}\dot{\rho}^2 + \frac{1}{2} \lambda_0^2 \rho^2
+ \sum_{j=1}^{N} \left[ \frac{1}{2} \dot{y}_j^2 + \frac{1}{2}
\lambda_j^2 y_j^2 \right]
\end{equation}

The eigenvalues $\lambda_0^2$ and $\lambda_j^2$ are expressible
in terms of the coupling constants of the system and the bath
implicitly as follows:

\begin{eqnarray}
\lambda_0^2 & = & \omega_0^2 / \left[1 - \frac{1}{\mu}\sum_{j=1}^N
\frac{c_j^2}{m_j \omega_j^2 (\lambda_0^2-\omega_j^2)} \right] \label{32}\\
\lambda_i^2 & = & \omega_0^2 / \left[1 - \frac{1}{\mu}\sum_{j=1}^N
\frac{c_j^2}{m_j \omega_j^2 (\lambda_i^2-\omega_j^2)}
\right],\;\;\; i=1,2...N \label{33}
\end{eqnarray}

where (\ref{32}) and (\ref{33}) correspond to normal mode
frequencies of the system and $i$-th bath oscillator respectively.

The transformation (\ref{30}) explicitly implies,

\begin{equation}\label{34}
q^{\prime}=u_{00}\; \rho + \sum_{j=1}^{N}u_{j0}\;y_j
\end{equation}

where the matrix elements $u_{00}$ and $u_{j0}$ can be expressed
as

\begin{eqnarray}
u_{00}^2 & = & 1 / \left[1 + \frac{1}{\mu}\sum_{j=1}^N
\frac{c_j^2}{m_j (\lambda_0^2-\omega_j^2)^2} \right] \label{35}\\
u_{i0}^2 & = & 1 / \left[1 + \frac{1}{\mu}\sum_{j=1}^N
\frac{c_j^2}{m_j (\lambda_i^2-\omega_j^2)^2} \right],\;\;\;
i=1,2...N \label{36}
\end{eqnarray}

For the present problem of dephasing it is significant to
consider the coupling between the system and the bath modes to be
weak. One can make use of the perturbation technique and
$\lambda_0^2$ and $\lambda_j^2$ are then expressible in simple
terms as \cite{pol},

\begin{eqnarray}
\lambda_0^2 & = & \omega_0^2 \left[1 - \frac{1}{\mu}\sum_{j=1}^N
\frac{c_j^2}{m_j\; \omega_j^2\; (\omega_j^2-\omega_0^2)} \right]+O(c_j^4) \nonumber\\
\omega_i^2 & = & \omega_j^2 \left[1 + \frac{c_j^2}{\mu\; m_j\;
\omega_j^2\; (\omega_j^2-\omega_0^2)}
\right]+O(c_j^4),\;\;\; j=1,2.....N \nonumber\\
u_{j0}& = &
-u_{0j}=\frac{c_j}{\sqrt{\mu\;m_j}\;(\omega_j^2-\omega_0^2)}+O(c_j^2),\;\;\;
j=1,2.....N\label{37}\\
u_{00}& = & 1+O(c_j^2)\nonumber\\
u_{ij}& = & 0+O(c_j^2),\;\;\;\;ij\ne0\nonumber
\end{eqnarray}

\section{{\bf quantum vibrational dephasing rate for a cubic oscillator}}

It has already been established \cite{ox1,ox2,ox3} that harmonic
oscillator model is not sufficient for a quantitative description
of vibrational dephasing rate. The essential paradigm for the
theory that has been used over the decades involves cubic
nonlinearity so that the potential assumes the following form,

\begin{equation}\label{38}
V(q)=\frac{1}{2}\mu\;\omega_0^2\;q^2 + \frac{1}{6}f\;q^3
\end{equation}

Here $f$ is a constant due to cubic nonlinearity. With Eq.(\ref{38})
the full Hamiltonian $H$[(\ref{25})] in normal co-ordinate is given by

\begin{equation}\label{39}
H=H_{NM}+k_{111}(u_{00}\; \rho + \sum_{j=1}^{N}u_{j0}\;y_j)^3+
3\mu B_2 u_{00}^2 k_{111}(u_{00}\; \rho +
\sum_{j=1}^{N}u_{j0}\;y_j)+\mu^{3/2}B_3u_{00}^3 k_{111}
\end{equation}

where we have used Oxtoby's notation, $k_{111}=f/{6\mu^{3/2}}$
and relations (\ref{27}) and (\ref{28}). Here the first term
denotes the normal mode Hamiltonian for the harmonic potential
and the second term corresponds to classical nonlinear part of
the potential. In addition to a constant shift of quantum origin
third term signifies the quantum corrections to system normal
mode where nonlinearity and quantum effects are entangled. In
what follows we show that this term provides a substantial
contribution to the vibrational dephasing rate.

The anharmonicity in the potential shifts the minimum and the
frequency of the system normal mode so that by applying the usual
condition

\begin{equation}\label{40}
\left(\frac{\partial H}{\partial \rho}\right)_{\rho_e}=0
\end{equation}

to obtain the instantaneous minimum of the potential, $\rho_e$,
we have

\begin{equation}\label{41}
\rho_e= \frac{1}{6k_{111}u_{000}^3}\left[
-(\lambda_0^2+6k_{111}u_{00}^2\sum_{j=1}^N
u_{j0}y_j)+(\lambda_0^4+12\lambda_0^2k_{111}u_{00}^2\sum_{j=1}^N
u_{j0}y_j-36 \mu B_2 u_{00}^6 k_{111}^2)^{1/2} \right]
\end{equation}

The instantaneous frequency is given by

\begin{eqnarray}
\lambda_0(t)& = &
\left(\frac{\partial^2H}{\partial\rho^2}\right)^{1/2}_{\rho_e}\nonumber\\
& = &
\lambda_0\left[1+\frac{12\;u_{00}^2\;k_{111}}{\lambda_0^2}\sum_{j=1}^N
u_{j0}\;y_j-\frac{36\;\mu\; B_2(t)\;u_{00}^6\;k_{111}^2}{\lambda_0^4}\right]^{1/4}\nonumber\\
& \simeq &
\lambda_0\left[1+\frac{3\;k_{111}}{\lambda_0^2}\sum_{j=1}^N
u_{j0}\;y_j-\frac{9\;\mu\;
B_2(t)\;k_{111}^2}{\lambda_0^4}\right]\label{42}
\end{eqnarray}

where we have used $u_{00}$ in the leading order. The instantaneous
frequency shift is therefore,

\begin{eqnarray}
\Delta \omega(t)& = & \lambda_0(t)-\lambda_0\nonumber\\
& = & \frac{3k_{111}}{\lambda_0}\sum_{j=1}^N u_{j0}y_j-\frac{9\mu
B_2(t)k_{111}^2}{\lambda_0^3}\label{43}
\end{eqnarray}

In the weak coupling limit the dephasing rate is expressed as,

\begin{equation}\label{44}
\kappa_{dep}=\int_0^\infty \langle \Delta \omega(t) \Delta
\omega(0)\rangle \;dt
\end{equation}

where the averaging is carried out over the thermally distributed
bath modes

\begin{equation}\label{45}
\langle \Delta \omega(t) \Delta
\omega(0)\rangle=\frac{9k_{111}^2}{\lambda_0^2}\sum_{j=1}^N
\frac{u_{j0}^2}{\lambda_j^2}\left[ \frac{1}{2} \hbar \lambda_j
\coth\left(\frac{\hbar\lambda_j}{2 k_B T}\right) \right]
\cos(\lambda_jt)+\frac{81\mu^2k_{111}^4}{\lambda_0^6}B_2(t)B_2(0)
\end{equation}

Here we have used the relations for the thermalized bath modes
\cite{hil,db1,skb,db2,db3,bk1,bk2,db4},

\begin{eqnarray}
\langle y_i(0) \rangle_{S} = \langle \dot y_i(0) \rangle_{S} =
\langle y_i(0) \dot y_i(0) \rangle_{S}=0\nonumber\\
\langle \dot y_i(0)^2 \rangle_{S} = \lambda_i^2 \langle y_i(0)^2
\rangle_{S} = \frac{1}{2} \hbar \lambda_i
\coth\left(\frac{\hbar\lambda_i}{2 k_B T}\right)\label{46}
\end{eqnarray}

The quantum dephasing rate is given by

\begin{equation}\label{47}
\kappa_{dep}=\frac{9k_{111}^2}{\lambda_0^2} \int_0^{\infty}dt\;
\sum_{j=1}^N \frac{u_{j0}^2}{\lambda_j^2}\left[ \frac{1}{2} \hbar
\lambda_j \coth\left(\frac{\hbar\lambda_j}{2 k_B T}\right) \right]
\cos(\lambda_jt)+\frac{81\mu^2k_{111}^4 B_2(0)}{\lambda_0^6}\int_0^\infty
dt\; B_2(t)
\end{equation}

Using (\ref{37}) for the expressions $\lambda_0^2$, $\lambda_j^2$
and $u_{j0}^2$ in (\ref{47}) we obtain in the weak coupling regime

\begin{eqnarray}
\kappa_{dep}& = &\frac{9k_{111}^2}{\omega_0^2} \int_0^{\infty}dt\;
\sum_{j=1}^N \frac{c_{j}^2}{\mu\; m_j\;
\omega_j^2}\frac{1}{(\omega_j^2-\omega_0^2)^2}\left[ \frac{1}{2}
\hbar \omega_j \coth\left(\frac{\hbar\omega_j}{2 k_B T}\right)
\right] \cos(\omega_jt)\nonumber\\
& + & \frac{81\mu^2k_{111}^4 B_2(0)}{\lambda_0^6}\int_0^\infty dt\;
B_2(t)\label{48}
\end{eqnarray}

Use of Eq.(\ref{22}) in the above expression and continuum limit
results in

\begin{eqnarray}
\kappa_{dep}& = &\frac{18k_{111}^2}{\pi \omega_0^2}
\int_0^{\infty}dt\int_0^\infty d\omega\; \frac{J(\omega)}{\omega}
\frac{1}{(\omega^2-\omega_0^2)^2}\left[ \frac{1}{2} \hbar \omega
\coth\left(\frac{\hbar\omega}{2 k_B T}\right)
\right] \cos(\omega t)\nonumber\\
& + & \frac{81\mu^2k_{111}^4 B_2(0)}{\lambda_0^6}\int_0^\infty dt\;
B_2(t)\label{49}
\end{eqnarray}

This is the general expression for quantum vibrational dephasing
rate. The essential content of this formulae in addition to the
usual first term obtained earlier by Levine \textit{et al}
\cite{pol} is the second term signifying the quantum contribution
to dephasing rate arising out of the nonlinearity of the system
potential. This term is independent of the quantum nature of the
thermal bath. An evaluation of this term requires the explicit
calculation of the integral over quantum correction term $B_2(t)$
which we pursue in the next section. Keeping in view of the fact
that $J(\omega)$ does not involve a specific choice of form for
density of bath modes, we find that the expression for the
dephasing rate as derived above is fairly general.

The above method is based on the normal mode Hamiltonian of Pollak
\cite{p1} adopted to a c-number description. An analysis of pure
dephasing of a nonlinear vibrational mode has been worked out
earlier to calculate non-Markovian line shape by Georgievskii and
Stuchebrukhov \cite{geo} using normal mode Hamiltonian treated by
thermodynamic Green's function approach. While the basis of
present calculation of dephasing rate is Eq.(\ref{44}), the
authors of Ref.5 have taken recourse to a different strategy to
calculate the line shape. The differences in formulations and
starting Hamiltonians (In Ref[5] a Leggett-Caldeira form of
Hamiltonian, \textit{i.e.}, Eq.(\ref{1}) without a counter term
has also been employed) notwithstanding, the effect of a quantum
contribution to dephasing width related to anharmonicity of the
oscillator has been calculated in both Ref[5] and present
analysis. The effect is due to the fact that the frequency of
fundamental transition of a quantum nonlinear oscillator differs
from harmonic frequency. To this end a continuation of the
present analysis to calculate the associated frequency shift is
instructive for comparison with those of others \cite{geo,pol}.
For this we return to the expression (\ref{42}) for instantaneous
frequency $\lambda_0(t)$, which after keeping terms upto
$k_{111}^2$ may be written as

\begin{equation}\label{n1}
\lambda_0(t) =
\lambda_0\left[1+\frac{3k_{111}}{\lambda_0^2}\sum_{j=1}^N
u_{j0}\;y_j(t)- \frac{27k_{111}^2}{2\lambda_0^4}\left(\sum_{j=1}^N
u_{j0}\;y_j(t)\right)^2 - \frac{9\mu
B_2(t)k_{111}^2}{\lambda_0^4}\right]
\end{equation}

where we have put $u_{00}$ to the leading order (unity). The time
average frequency $\overline{\lambda}_0$ is given by \cite{pol}

\begin{equation}\label{n2}
\overline{\lambda}_0=\lim_{\tau\rightarrow\infty} \frac{1}{\tau}
\int_0^{\tau} \lambda_0(t^{\prime})\; dt^{\prime}
\end{equation}

Putting (\ref{n1}) in (\ref{n2}) we obtain

\begin{equation}\label{n3}
\overline{\lambda}_0=\lambda_0\left[1 -
\frac{27\;k_{111}^2}{2\;\lambda_0^4}\sum_{j=1}^N
\frac{u_{j0}^2}{\lambda_j^2}\left( \frac{1}{2}\hbar\lambda_j
\right)\coth\left(\frac{\hbar\lambda_j}{2k_B T}\right) -
\frac{9\;\mu\; \overline{B}_2\;k_{111}^2}{\lambda_0^4}\right]
\end{equation}

where we have used (\ref{46}) and $\overline{B}_2 \;(=\omega_c
\int_0^{1/\omega_c} B_2(t^\prime) dt^\prime$ , $\omega_c$ being
the cutoff frequency) is given by
$\frac{\hbar\omega_c\gamma}{8\mu\omega_0^3}$ (the explicit form
of $B_2(t)$ is calculated in the next section). Furthermore with
the replacement of $u_{j0}$ using Eq.(\ref{37}) and then using
Eq.(\ref{22}) in the continuum limit we obtain

\begin{equation}\label{n4}
\overline{\lambda}_0=\lambda_0 -
\frac{27\;k_{111}^2\hbar}{2\;\pi\lambda_0^3\omega_0^4}
\int_0^{\omega_c} J(\omega) \coth\left(\frac{\hbar\omega}{2k_B T
}\right) d\omega - \frac{9 k_{111}^2 \hbar \omega_c \gamma}{8
\lambda_0^3 \omega_0^3}
\end{equation}

Since $\lambda_0^2$ is given by \cite{pol},

\begin{equation}\label{n5}
\lambda_0^2=\omega_0^2+\frac{2}{\pi}\int_0^{\omega_c}
\frac{J(\omega)}{\omega} d\omega
\end{equation}

one may use (\ref{58}) to obtain

\begin{equation}\label{n6}
\lambda_0 = \omega_0 \left[1+\frac{\gamma
\omega_c}{4\pi\omega_0^2}\right]
\end{equation}

The frequency shift $\Delta\omega_0$ is then derived from
Eq.(\ref{n4}) and Eq.(\ref{n6}) as

\begin{eqnarray}
\Delta\omega_0 & = & \overline{\lambda}_0 - \omega_0\nonumber\\
& = & \frac{\gamma \omega_c}{4\pi\omega_0^2}-\frac{27\;\gamma
k_{111}^2}{8\;\pi\omega_0^7} \int_0^{\omega_c} \hbar \omega
\coth\left(\frac{\hbar\omega}{2k_B T }\right) d\omega - \frac{9
k_{111}^2 \hbar \omega_c \gamma}{8 \omega_0^6}\label{n7}
\end{eqnarray}

It is important to note that while in addition to the first two
terms corresponding to treatment of Levine, Shapiro and Pollak
\cite{pol} the last one refers to temperature independent
anharmonic quantum contribution proportional to $k_{111}^2$
responsible to the frequency shift noted earlier in Ref.[5]. We
mention in passing that the presence and absence of the counter
term in the Hamiltonian may cause a significant difference in
frequency shift with respect to direction towards blue or red
region \cite{geo}.

\section{{\bf calculation of the quantum correction due to nonlinearity of
the system potential}}

It has already been pointed out that a leading order quantum
correction due to nonlinearity of the potential of the system
provides an important contribution over and above the usual
expression for dephasing rate. To calculate this term explicitly
we now return to the operator equation (\ref{2}) and use
(\ref{18}) and (\ref{19}) to obtain

\begin{equation}\label{50}
\mu \;\delta \dot {\hat q}=\delta \hat p
\end{equation}

\begin{equation}\label{51}
\delta \dot {\hat p} + \int_0^t \gamma(t-t^\prime) \;\delta \hat
p(t^\prime)\; dt^\prime + V^{\prime \prime} (q) \;\delta \hat q +
\sum_{n\geq 2}\frac{1}{2} \;V^{(n+1)}(q)\; (\delta \hat
q^n-\langle \delta \hat q^n\rangle) =\hat{F}(t)-f(t)
\end{equation}

We then perform a quantum mechanical averaging over bath states
with $\prod_{i=1}^N \{ | \alpha_{i}(0) \rangle \}$  to get rid of
the term $\hat F(t)-f(t)$. The Eqs.(\ref{50}) and (\ref{51}) along
with (\ref{15}) and (\ref{16}) form the key element for
calculation of the quantum mechanical correction. Considering the
friction kernel $\gamma(t)$ to be arbitrary (but decaying) we may
calculate the leading order quantum correction for the harmonic
mode for which higher derivatives of $V(q)$ in (\ref{51}) vanish.
Now Eq.(\ref{51}) becomes

\begin{equation}\label{52}
\delta \dot {\hat p}(t)=- \int_0^t \gamma(t-t^\prime) \;\delta
\hat p(t^\prime)\; dt^\prime - \mu \;\omega_0^2\delta \hat q(t)
\end{equation}

where $\mu \;\omega_0^2=V^{\prime\prime}(q)$ corresponding to the
harmonic mode. The above equations (\ref{50}) and (\ref{52}) can
then be solved by Laplace transformation technique to obtain

\begin{equation}\label{53}
\delta {\hat p}(t)=\frac{1}{\mu}\;\delta \hat p(0)\; C_v(t) +
\delta {\hat q}(0)\; C_q(t)
\end{equation}

where

\begin{equation}\label{54}
C_v(t)=L^{-1} \left[ \frac{1}{s^2+s\; \widetilde{\gamma}(s) +
\omega_0^2} \right]
\end{equation}

and

\begin{equation}\label{55}
C_q(t) = 1 - \omega_0^2 \int_0^t C_v(t^\prime) \;dt^\prime
\end{equation}

and $\widetilde{\gamma}(s)$ is the Laplace transform of
$\gamma(t)$ defined as $\widetilde{\gamma}(s)=\int_0^\infty
\gamma(t) e^{-s t} dt$. After squaring and quantum mechanical
averaging Eq.(\ref{53}) yields

\begin{equation}\label{56}
\langle \delta \hat q^2(t) \rangle = \frac{1}{\mu^2}\;\langle
\delta \hat p^2(0)\; \rangle C_v^2(t) + \langle \delta \hat
q^2(0)\; \rangle C_q^2(t) + \frac{1}{\mu}\;C_v(t)\; C_q(t) \langle
\delta \hat p(0)\; \delta \hat q(0)+\delta \hat q(0)\; \delta
\hat p(0) \rangle
\end{equation}

For a minimum uncertainty state we chose \cite{sm}

\begin{equation}\label{57}
\langle \delta \hat p^2(0)\;
\rangle=\frac{\mu\hbar\omega_0}{2},\;\;\;\langle \delta \hat
q^2(0)\; \rangle=\frac{\hbar}{2\mu\omega_0}\;\;\;and\;\;\;\langle
\delta \hat p(0)\; \delta \hat q(0)+\delta \hat q(0)\; \delta
\hat p(0) \rangle=0
\end{equation}

Furthermore we assume the form of the spectral density function,
$J(\omega)$, as,

\begin{equation}\label{58}
J(\omega)=\frac{1}{4}\; \gamma\;\omega
\end{equation}

where $\gamma$ is the static dissipation constant.

Using Eqs.(\ref{4}), (\ref{22}) and (\ref{58}) in the continuum
limit we have

\begin{equation}\label{59}
\gamma(t)=\frac{1}{2}\;\gamma\;\delta(t)
\end{equation}

Laplace transform results in

\begin{equation}\label{60}
\gamma(s)=\gamma_1
\end{equation}

where $\gamma_1=\gamma/2$.

Now with the form of $\gamma(s)$ as given by Eq.(\ref{60}) the
relaxation functions $C_v(t)$ and $C_q(t)$ become

\begin{eqnarray}
C_v(t)& = &\frac{1}{2\omega_1}\left[ e^{-s_1 t}-e^{-s_2 t}
\right]\label{61}\\
C_q(t)& = &\frac{\omega^2}{2\omega_1}\left[ \frac{1}{s_1}\;e^{-s_1
t}-\frac{1}{s_2}\;e^{-s_2 t} \right]\label{62}
\end{eqnarray}

where,

\begin{equation}
\omega_1 =
\left[\frac{\gamma_1^2}{4}-\omega_0^2\right]^{1/2},\;\;\; s_1 =
\frac{\gamma_1}{2}-\omega_1,\;\;\; s_2
=\frac{\gamma_1}{2}+\omega_1\label{63}
\end{equation}

Making use of Eq.(\ref{57}), (\ref{61}) and (\ref{62}) in
(\ref{57}) we obtain quantum correction term $B_2(t)(=\langle
\delta \hat q^2(t) \rangle)$ as

\begin{equation}\label{64}
B_2(t)=\frac{\hbar\omega_0}{8\mu
\omega_1^2}\left[\left(1+\frac{\omega_0^2}{s_1^2}\right)e^{-2s_1t}+
\left(1+\frac{\omega_0^2}{s_2^2}\right)e^{-2s_2t}-4e^{-\gamma_1t}\right]
\end{equation}

The above term can be utilized in the integral of the second term
in Eq.(\ref{49}) for its explicit evaluation to find out the
dependence of the system parameters on the dephasing rate
analytically. For better accuracy the systematic corrections to
higher order can be worked out as discussed in detail in
Refs.[18-24].

\section{{\bf vibrational dephasing rate; comparison with experiments
and discussion}}

Having obtained the explicit expression for the leading order
contribution $B_2(t)$ from Eq.{\ref{64}} we are now in a position
to write down the total quantum vibrational dephasing rate. To
this end we make use of Eq.{\ref{64}} in the second term and
Eq.(\ref{58}) in first term of the expression (\ref{49}) and
obtain, after some algebra,

\begin{equation}\label{65}
\kappa_{dep}=\kappa_1+\kappa_2
\end{equation}

with

\begin{equation}
\kappa_1 = \frac{9\;k_{111}^2\;\gamma}{4\; \pi\; \omega_0^2}
\int_0^{\infty}dt\int_0^\infty d\omega\; \frac{\hbar \omega
}{(\omega^2-\omega_0^2)^2} \coth\left(\frac{\hbar\omega}{2 k_B
T}\right) \cos(\omega t)\label{66}
\end{equation}

and

\begin{equation}
\kappa_2 = \frac{81\;\hbar^2\; k_{111}^4
\gamma}{16\;\omega^{10}}\label{67}
\end{equation}

The vibrational dephasing time can be defined as

\begin{equation}\label{68}
\tau_v=\frac{1}{\kappa_{dep}}
\end{equation}

The expression (\ref{65}) is the central result of this paper. We
already pointed out that $\kappa_2$ is a new contribution of
quantum origin due to nonlinearity of the system potential.
$\kappa_1$ in the limit $k_B T\gg \hbar \omega_0$ is the standard
well known expression for the classical dephasing rate. It is
important to note that $\kappa_1$ incorporates quantum effect due
to heat bath only. Although both $\kappa_1$ and $\kappa_2$ are
dependent on nonlinearity, $\kappa_2$ vanishes in the classical
limit and is also independent of temperature. The temperature
dependence of the dephasing rate is due to the first term
$\kappa_1$ of Eq.(\ref{66}). It is important to note that at very
low temperature as the integrand $\frac{1}{2} \hbar\omega
\coth(\hbar\omega/2k_BT)$ in $\kappa_1$ reduces to $\frac{1}{2}
\hbar\omega$, the temperature independent vacuum limit, one
observes that the dephasing caused by anharmonicity of the
vibrational mode does not vanish even at absolute zero because of
the contributions of these two terms. This aspect of temperature
independence of the width of the transition from the ground state
had been noted earlier in Ref.[5]. The origin of the temperature
independence in $\kappa_2$ and the nature of dephasing may be
traced to the second term in Eq.({\ref{45}}) which results from
the third term of the normal mode Hamiltonian Eq.(\ref{39}). This
term contains the quantum contribution to nonlinear potential
explicitly calculated in terms of $B_2(t)$ in Sec.IV. This lowest
order quantum fluctuation (or uncertainty) is independent of the
quantum character of the heat bath and also temperature and causes
frequency fluctuation leading to dephasing and therefore a
homogeneous broadening of the transition similar to natural
linewidth. In order to assess the relative contribution of the
two terms in the total dephasing rate we estimate the numerical
magnitude of these two quantities as well as the dephasing time
for three diatomic molecules, $N_2$, $O_2$ and $CO$ and compare
them with experimental results obtained from either picosecond
pump-probe technique \cite{lau} or from Raman linewidth
measurement of liquids using interferometric techniques
\cite{wrl,sco,srj,clo1,clo2}. We have also studied the rate as a
function of temperature away from critical point or triple point
of these liquids and compared with experiments.

The values of the parameters essential for calculation of the
dephasing rate using formulae (\ref{65}) are given in the Table
1. Apart from mass $\mu$, frequency of the fundamental
$\omega_0$, size $r$, at a temperature $T$ two sensitive
parameters are the static friction due to the liquid, $\gamma$,
and the anharmonic force constant $k_{111}$. Although use of
local viscosity which formally takes into account of wave vector
dependence of the viscosity for the size of the probe has been
advocated for calculation of friction, we confine ourselves to
standard Stoke's expression ($\gamma=6\pi\eta r/\mu$, $\eta$ being
viscosity coefficient of the liquid). For diatomics we determine
$k_{111}$ from the spectroscopic constants $\alpha_e$ and $B_e$
using \cite{ox4,her} $k_{111}=-\frac{\hbar \omega_0^2}{4
\mu^{3/2} B_e r^3}(1+\frac{\alpha_e\omega_0}{6B_e^2})$. The
integrals in (\ref{66}) are calculated numerically for all the
three cases. The magnitudes of $\kappa_1$ and $\kappa_2$ are
shown separately in the table along with the percentage
contribution of $\kappa_2$ in the total dephasing rate
$\kappa_{dep}$. Three pertinent points are to be noted. First, it
is well known that classical dephasing rate (\textit{i.e.}
$\kappa_1$ in the classical limit) is higher than the
corresponding quantum rate $\kappa_1$. This is somewhat
reminiscent of the quantum supression of classical values of rate
co-efficient for the thermally activated processes for shallow
cubic potentials \cite{gri}. Second, it is evident that just as
in the classical theory, anharmonicity contributes significantly
to the total quantum dephasing rate. In case of $N_2$ it is as
large as $\sim 26.5$ percent of total dephasing rate. The
dephasing time thus calculated corresponds fairly to that
obtained from experiments. Third, the quantum effects due to heat
bath appears quite significantly through the integrals in
$\kappa_1$. This is because, the frequency dependence of the
integrand is quite sharp at around $\omega_0$ as a result of the
the frequency denominator $(\omega^2-\omega_0^2)$ and
$\hbar\;\omega_0\gg k_B\;T$ for these molecules. Therefore the
diatomic oscillator behaves more closely as a quantum oscillator
rather than its classical counterpart.

The temperature dependence of the dephasing rate according to
Eq.(\ref{65}) has been compared for the set of parameter values
mentioned in the Table 1 to that obtained from experiments
\cite{clo1,clo2} in Fig.1 for the molecules $N_2$, $O_2$ and
$CO$. While the experimental results span a wide range of
temperatures covering triple point and critical point, we confine
ourselves in the region away from the points of phase transition,
since the present theory is outside the scope of phase
transition. We find a linear dependence of dephasing rate on
temperature which is a fair agreement with experimental
observations \cite{clo1,clo2}.

\section{{\bf conclusion}}

Based on a quantum Langevin equation and the corresponding
Hamiltonian formulation within a c-number description, we have
calculated quantum vibrational dephasing rate for a cubic
oscillator system using first order perturbation technique and
compared the results with experiments. It is shown that the
vibrational dephasing rate comprises of two terms. The first term
is the standard expression obtained earlier by several workers
using independent approaches and is reduces to its classical
limit as $k_B\;T\gg\hbar\;\omega_0$. This term is responsible for
finite temperature dependence of the dephasing rate. The second
term, an essential content and offshoot of the present theory, is
a leading quantum correction due to nonlinearity of the system
potential which contributes quite effectively to the total
dephasing rate. Keeping in view of this observation, we believe
that this quantum correction  term is likely to play a significant
role in triatomic and polyatomic molecules as well.

{\bf Acknowledgement}\\
We thank Prof. B. Bagchi for interesting discussions. The authors
are indebted to the Council of Scientific and Industrial Research
for partial financial support under Grant No. 01/(1740)/02/EMR-II.

\newpage
\begin{center}
{\bf Figure Captions}
\end{center}

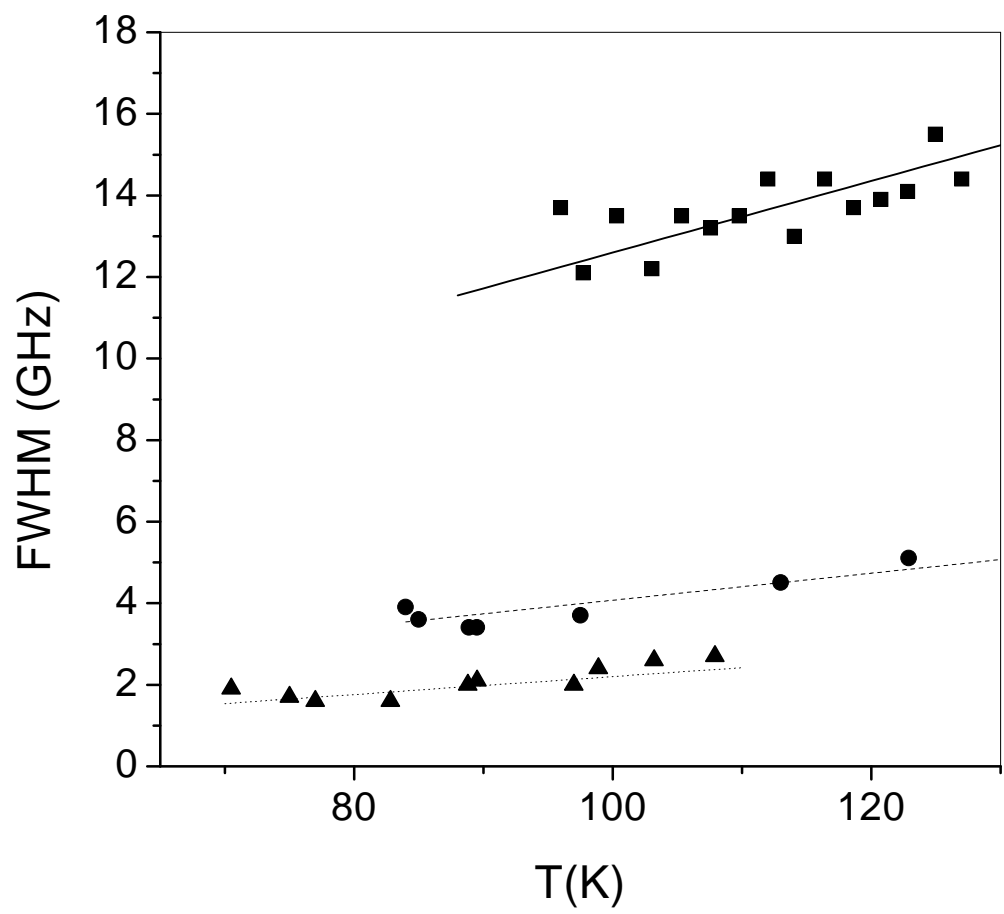
Fig.1: Variation of vibrational dephasing rate (FWHM) with
temperature (T) for the set of parameter values mentioned in the
Table 1 and comparison with experiments for three different
diatomics $N_2$ ($\blacktriangle$ experiment \cite{clo1} ; dotted
line, theory), $O_2$ ($\bullet$ experiment \cite{clo1}; dashed
line, theory) and $CO$ ($\blacksquare$ experiment \cite{clo2};
bold line, theory).

\end{document}